

\input harvmac
\def\t{\theta}
\def\g{\gamma}
\def\p{\prime}
\def\<{\langle}
\def\>{\rangle}
\lref\miwa{ M. Jimbo, R. Kedem, T. Kojima, H. Konno and T. Miwa,
"XXZ chain with a boundary",  hep-th/9411112.}
\lref\Kor{V.E.Korepin, Theor. Math. Phys. 41 (1979) 953 \semi
H.Bergknoff, H.Thacker, Phys.Rev. D19 (1979) 3666.}
\lref\Gaud{M.Gaudin, "La Fonction d'Onde de Bethe", Masson (Paris).}
\lref\Wch{F.Woynarovich, Phys.Lett. A108 (1985) 401.}
\lref\ABBBQ{F.Alcaraz, M.Barber, M.Batchelor, R.Baxter, G.Quispel,
J.Phys.  A20 (1987) 6397.}
\lref\SSW{H.Saleur, S.Skorik, N.P.Warner "The boundary sine-Gordon
theory:
classical and semi-classical analysis", preprint USC-94-013,
hep-th/9408004.}
\lref\GZ{S.Ghoshal, A.Zamolodchikov, Int.J.Mod.Phys.  A9 (1994)
3841.}
\lref\DdVg{C. Destri, H.J. de Vega, "Unified approach to TBA
and finite size
corrections for lattice models and field theories",
preprint LPTHE 94-28.}
\lref\Ghosh{S.Ghoshal, Int.J.Mod.Phys. A9 (1994) 4801.}
\lref\FS{P.Fendley, H.Saleur, Nucl. Phys. B428 (1994) 681.}
\lref\Mc{A.MacIntyre, "Integrable boundary conditions for classical
sine-Gordon theory", preprint DTP/94-39, hep-th/9410026.}
\lref\ZZ{A.B.Zamolodchikov, Al.B.Zamolodchikov, Ann. Phys.  120
(1980)
253.}
\lref\skly{E.K. Sklyanin, J. Phys. A21 (1988) 2375.}
\lref\DDV{C.Destri, H. de Vega, J.Phys. A22 (1989) 1329.}
\lref\RS{N.Reshetikhin, H.Saleur,  Nucl. Phys. B419 (1994) 507.}
\lref\BdVV{O.Babelon, H.J. de Vega, C.M.Viallet, Nucl. Phys. B220
(1983) 13.}
\lref\fk{A. Fring and R. K\"oberle, Nucl.Phys. B419 (1994) 647; {\it
ibid}
B421 (1994) 1592.}
\lref\Corr{E.Corrigan, P.E.Dorey, R.H.Rietdijk, R.Sasaki, Phys. Lett.
B333
(1994) 83.}
\lref\IK{T.Inami, H.Konno, "Integrable XYZ Spin Chain with
Boundaries", preprint YITP/K-1084, hep-th/9409138, to appear in
J.Phys. A Lett.}
\lref\sasaki{R. Sasaki, ``Reflection Bootstrap Equations for Toda
Field
Theory'',  preprint YITP/U-93-33, hep-th/9311027.}
\lref\YY{C.N.Yang and C.P.Yang, Phys.Rev 150 (1966) 327.}
\lref\Skl{E.K.Sklyanin, Funct. Anal. and Appl. 21 (1987) 164.}

\noblackbox
\Title{\vbox{\baselineskip12pt
\hbox{USC-95-01}\hbox{hep-th/9502011}}}
{\vbox{\centerline{Boundary bound states and boundary
bootstrap}\vskip4pt
\centerline{ in
the sine-Gordon model}\vskip4pt
\centerline{ with Dirichlet boundary conditions}}}
\centerline{Sergei Skorik$^\dagger$ and Hubert Saleur$^{\ddagger*}$}
\vskip4pt
\centerline{\it $^\dagger$Department of Physics, University of
Southern
California}
\centerline{\it Los Angeles CA 90089-0484}
\vskip2pt
\centerline{\it $^\ddagger$Department of Physics and Department of
Mathematics}
\centerline{\it University of Southern California}
\centerline{\it Los Angeles CA 90089}
\vskip.3in

We present a complete study of boundary bound states
and related boundary S-matrices for the sine-Gordon model
with Dirichlet boundary conditions.  Our approach is based partly
on the bootstrap procedure, and partly on  the
explicit solution of the inhomogeneous XXZ model with boundary magnetic field
and of the
boundary Thirring model. We identify boundary bound states with new ``boundary
strings'' in the Bethe ansatz. The boundary energy is also computed.

\bigskip
\bigskip\bigskip\bigskip
\noindent $^*$ Packard Fellow
\Date{February, 1995}

\newsec{Introduction.}

The sine-Gordon model with a boundary interaction preserving
integrability (which we shall call the boundary sine-Gordon model) is
of theoretical as well as practical interest. In particular, it
exhibits relations with the theory of Jack symmetric functions
\ref\JACK{P. Fendley, F. Lesage, H. Saleur, ``Solving 1D plasmas and
2D boundary problems using Jack polynomials and functional
relations'', preprint USC-94-16, hep-th/9409176.} and has appplications to
dissipative
quantum mechanics \ref\DQM{F. Guinea, V. Hakim, A. Muramatsu, Phys.
Rev. Lett. 54 (1985) 263.} and impurity problems in
1D strongly correlated electron gas \ref\KF{C. Kane, M. Fisher, Phys.
Rev. B46 (1992) 15233.}.

In the seminal work \GZ\ it appeared clearly that this problem
presents an extremely  rich structure of boundary bound states, which
was partly explored  in \Ghosh. Our first purpose here is to
study this structure further in the particular case of Dirichlet
boundary conditions, that is
the model
\eqn\SG{{\cal L}_{SG}={1\over 2} \int_0^{\infty}\left[(\partial_t\varphi)^2-
(\partial_x\varphi)^2 + {m_0^2\over\beta^2}\cos\beta\varphi\right]dx}
with a fixed value of the field at the boundary: $\varphi(x=0,
t)=\varphi_0$.

Also, the consideration of boundary problems poses interesting
challenges from the point of view of lattice models, here lattice
regularizations of \SG. In \FS\ and also in \ref\Lucaetal{M.T.
Grisaru, L. Mezincescu, R. Nepomechie, ``Direct calculation
of the boundary S matrix for the open Heisenberg chain'', hep-th/9407089.} it
was shown
in particular how to derive the S-matrices of \GZ\ from the Bethe
ansatz. Our second purpose is to complete these studies by
investigating which new types of strings correspond to boundary bound
states, and by deriving as well the set of S-matrices necessary to
close the bootstrap.
Observe that   lattice regularizations
are useful to define what one means
by putting a bound state
at the boundary. Indeed, some bound states have no
straightforward interpretation, and although they are easy to study
formally using the Yang Baxter equation and the bootstrap,
their meaning in the field theory is unclear.

In section 2 we consider the bootstrap problem directly in the
continuum theory. We identify boundary bound states and  we compute the related
boundary $S$ matrices.
In section 3 we write the Bethe ansatz equations for
the inhomogeneous six-vertex model with boundary magnetic field,
which is believed \FS\ to be a regularization of \SG. We  show that
these equations are also the bare equations for the Thirring model
with $U(1)$-preserving boundary interaction,
which is the fermionized version of \SG. In section 4 we discuss in
details new solutions (``boundary strings'') to the Bethe ansatz
equations made possible by the appearance of boundary terms. In
section 5 we study the physical properties of the model, in
particular the masses and S-matrices corresponding to these boundary
strings, and we partially complete the identification with the bootstrap
results of section 2. Several remarks, in particular formula for the
boundary energy of the
boundary sine-Gordon model, are collected in the conclusion.

\newsec{Boundary bootstrap results.}

\subsec{Solving the boundary bootstrap equations.}

The S-matrices for the scattering of a soliton  ($P^+$)
and an  anti-soliton ($P^-$) on the ground state
$|0\>_B$ of the sine-Gordon model with  Dirichlet
boundary conditions  \SG\ were obtained in \GZ:
\eqn\smatr{P^{\pm}(\t)= \cos(\xi\pm\lambda u)R_0(u)R_1(u,\xi),}
where $\t=iu$ is the rapidity, $\xi={4\pi\over\beta}\varphi_0$ and
$\lambda={8\pi\over\beta^2}-1$. The explicit form of $R_0, R_1$ is
rather cumbersome and can be found
in \GZ.
Since the theory is invariant under the simultaneous transformations
$\xi\rightarrow
-\xi$,
and soliton$\rightarrow$anti-soliton , we choose hereafter
$\xi$ to be a generic number in the interval $0<\xi<4\pi^2/\beta^2$
(about the value of the upper bound see the discussion in \SSW).

The function $R_0$ contains poles in the physical strip
$0<{\rm Im}\theta<\pi/2$ located at $u={n\pi\over
2\lambda},n=1,2,\ldots<\lambda$. These poles arise because of the
corresponding breather pole in the soliton-antisoliton bulk scattering,
and should not be interpreted as boundary bound states \GZ.

When $\xi>\pi/2$, the function $P^+(\theta)$ has additional poles in the
physical
strip
, located at  $u=v_n$ with :
\eqn\poles{0<v_n={\xi\over\lambda}-{2n+1\over 2\lambda}\pi<{\pi\over
2},}
$(n=0,1,2,\ldots)$ corresponding to  a first  set of boundary bound
states which we denote by $|\beta_n\>$, with
masses
\eqn\addi{m_n=m\cos v_n=m\cos\left({\xi\over\lambda}-{2n+1\over
2\lambda}\pi\right),}
where $m$ is the soliton mass. These bound states are easy to interpret
\GZ,\SSW.
For $0<\varphi_0<{\pi\over\beta}$ the ground state of the theory is
characterized by the asymptotic behaviour $\varphi\to 0$ as $x\to\infty$, but
other states, whose energy differs from the ground state by a boundary
 term only,
can be obtained with $\varphi\to \hbox{ a multiple of }{2\pi \over\beta}$
as $x\to\infty$.
Since the $\beta_n$ appear as bound states for soliton scattering,
they all have the same topological charge as the soliton, which
we take equal to unity by convention, so they are all associated with the same
classical solution,   a
 soliton
sitting next to the boundary and performing a motion
periodic in time ("breathing"), with $\varphi(x=0)=\varphi_0$ and
$\varphi\to{2\pi\over\beta}$ as $x\to\infty$ \SSW.

To deduce the scattering matrices on the boundary bound states we use
the "boundary bootstrap equations" as given in \GZ. We assume that
these S-matrices are diagonal, which is true if all the boundary
bound states
have different energies.  In this case the bootstrap equations read:
\eqn\BBE{R^b_{\beta}(\theta)=\sum_{c,d}R^d_{\alpha}(\theta)
S_{cd}^{ab}(\theta+iv^{\beta}_{\alpha a})
S_{ba}^{dc}(\theta-iv^{\beta}_{\alpha a}).}
These equations allow us to find the scattering matrix of any
particle
$b$  on
the boundary bound state $\beta$ provided that  the latter appears as
a virtual state
in the scattering  of the particle  $a$    on the boundary state
$\alpha$. The masses of the corresponding boundary states
are related through
\eqn\mass{m_{\beta}=m_{\alpha}+m_a\cos v^{\beta}_{\alpha a},}
where $iv^{\beta}_{\alpha a}$ denotes the position of the pole,
corresponding
to  the bound state $\beta$.

Let $\beta_n$ stand for the $n$-th boundary bound state
corresponding to
the pole $v_n$  in $P^+$ \poles. Then \BBE\ gives:
\eqn\EQI{P^+_{\beta_n}(\theta)=P^+(\theta)a(\theta-iv_n)a(\theta+iv_n)
,}
\eqn\EQII{P^-_{\beta_n}(\theta)=b(\theta-iv_n)b(\theta+iv_n)P^-(\theta
)
+ c(\theta-iv_n)c(\theta+iv_n)P^+(\theta),}
where the well known bulk S-matrix elements
$a(\theta)=S_{++}^{++}=S_{--}^{--}$
(kink-kink scattering),
 $b(\theta)=S_{+-}^{+-}=S_{-+}^{-+}$  (kink-anti-kink transmission)
and $c(\theta)=S_{+-}^{-+}=S_{-+}^{+-}$ (kink-anti-kink reflection)
can be  found in \ZZ.

It is easy to  check that the matrix elements \EQI-\EQII\ satisfy
general
requirements  for the boundary S-matrices, such as boundary unitarity
and boundary crossing-symmetry conditions \GZ, e.g.
$$ P^-_{\beta_n}({i\pi\over
2}-\theta)=b(2\theta)P^+_{\beta_n}({i\pi\over 2}+
\theta) + c(2\theta)P^-_{\beta_n}({i\pi\over 2}+\theta) ,$$
$$ P^{\pm}_{\beta_n}(\theta)P^{\pm}_{\beta_n}(-\theta)=1.$$
Finally  we obtain from \EQI-\EQII\ by direct calculation:
\eqn\BYBE{P^+_{\beta_n}(\theta)={\cos(\xi-\lambda\pi-i\lambda\theta)
\over\cos(\xi-\lambda\pi+i\lambda\theta)}P^-_{\beta_n}(\theta).}
Hence the boundary Yang Baxter equation  is satisfied since the ratio of the
above two
amplitudes
has a form similar to \smatr\ with $\xi\rightarrow\xi-\lambda\pi$,
$\xi$ being a free parameter.

The analytic structure of $P^{\pm}_{\beta_n}(\theta)$ is as follows.
The function $P^+_{\beta_n}(\theta)$ has
simple poles in the physical strip located at  $u={\xi\over\lambda}
+{2N+1\over 2\lambda}\pi$ ,  $N=0,1,2...$, and at $u=v_n$.
It has
double poles at  $u=iv_n+i{k\pi\over\lambda}$, $k=1,2,...n$.
The function $P^-_{\beta_n}(\theta)$  possesses
in the physical strip the same singularities as
$P^+_{\beta_n}(\theta)$
plus the set of simple poles at $u=iw_N$ with
\eqn\newpoles{w_{N}=\pi - {\xi\over\lambda} - {2N-1
\over 2\lambda}\pi, \qquad \lambda+{1\over 2}-{\xi\over\pi}
>N>{\lambda+1\over 2} - {\xi\over\pi}. }

Interpreting  these poles in terms of boundary bound states
requires some care. First, due to the relation \BBE, one sees that if $\beta$
appears as a boundary bound state for scattering of $a$ on $\alpha$, then
the poles of the amplitude for scattering of $b$ on $\alpha$ are also
in general  poles of the amplitude for scattering of $b$ on $\beta$. It seems
unlikely that these poles correspond to new bound states, although in our case
they would have a natural physical meaning, for example one could try to
associate them with classical solutions where $\varphi\to{4\pi\over\beta}$ as
$x\to\infty$. Indeed there are strong constraints coming
from statistics that we should not forget. For instance at the
free fermion point $\beta^2=4\pi$, there is a bound state $\beta_1$, but
although $P^+_{\beta_1}$ has again a pole at $\beta_1$, the state of mass
$2m_{\beta_1}$ is not allowed from Pauli exclusion principle, as can easily
be checked on the direct solution of the model (see below section 3.3).
Therefore we take the point
of view that the poles already present in the scattering on
 an ``empty boundary'' are ``redundant''.
The only poles we interpret as  related to
 new boundary bound states are \newpoles (the additional poles
 in $P_{\beta_n}^+$ are related to them by crossing). We denote these
boundary bound states $|\delta_{n,N}\>$, and their masses,
according to \mass\ and \poles, are given by
\eqn\massnew{\eqalign{m_{n,N}&=m(\cos v_n + \cos w_{N})
=m\cos\left({\xi\over\lambda}-{2n+1\over 2\lambda}\pi\right)-
m\cos\left({\xi\over\lambda}+{2N-1\over2\lambda}\pi\right)\cr
&=m^b_{N+n}\sin\left({\xi\over\lambda}+{N-n-1\over
2\lambda}\pi\right),\cr}}
where $m^b_p=2m\sin\left({p\pi\over 2\lambda}\right)$ is the mass of
the $p$-th breather, $p=1,2,\ldots<\lambda$.

To understand
the physical meaning of these new boundary bound states it is helpful
to consider the semi-classical limit $\lambda\to\infty$ of the
sine-Gordon
model.  As discussed above,
the boundary bound state $\beta_n$, corresponding to \poles, are
associated to solutions where a soliton is sitting next to the boundary and
``breathing''. An
incoming anti-soliton can couple to this soliton, and
together
they form  a breather sitting next to the boundary and performing again some
(rather complicated) motion periodic in time\foot{To compute this
solution explicitely requires using a  bulk five-soliton
configuration \SSW, an expression which is very cumbersome.}. The quantization
of this solution shoud lead to $|\delta_{n,N}\>$. The
topological charge of the states $|\delta_{n,N}\>$ is equal to 0 in our units,
or, equivalently, to the charge of a free breather in the theory \SG.

One can in principle continue to solve the bootstrap equations \BBE\
recursively. For example, for the scattering of solitons or
antisolitons on the boundary bound
states
$|\delta_{n,N}\>$ \newpoles\ one obtains the following S-matrices:
\eqn\FIN{P^-_{\delta_{n,N}}(\t)=P^-_{\beta_n}(\t)a(\t-iw_N)a(\t+iw_N),}
\eqn\FINI{P^+_{\delta_{n,N}}={\cos(\xi-i\lambda\t)\over\cos(\xi+i\lambda\t)}
P^-_{\delta_{n,N}}.}
$P^-_{\delta_{n,N}}$ has only one simple pole in the physical strip at
$u=w_N$,  while
$P^+_{\delta_{n,N}}$ has also simple poles at $u=v_k$, $k=n+1,
n+2,...,[{\xi\over
\lambda}-{1\over 2}]$. According to  the discussion below \newpoles, we
do not consider these poles as associated with new boundary bound states.
Therefore, the boundary bootstrap is closed
for
solitons and antisolitons in the sense
that  further recursion will not generate new boundary bound
states.

So far we have obtained  two sets of boundary bound states \addi\ and
\massnew\  by
considering all the poles in the physical strip of  amplitudes for
scattering a soliton (resp. anti-soliton) on a boundary with
 or without a boundary bound state.
Of course we should also consider the scattering of breathers off the
boundary.
The scattering of breathers off an ``empty'' boundary was studied in \Ghosh,
and we
refer the reader to this work for the explicit boundary S-matrices. By
interpreting  the poles
of the amplitudes in \Ghosh\  as boundary bound states, we find a
spectrum of masses that  look like
\massnew\ but with a slightly different range of parameters. Considering then
scattering of breathers off a boundary with a bound state does not give rise to
any new poles beside \poles\ and \newpoles, with in the latter case an extended
range of values of $N$ (for simplicity we do not
give the relevant boundary S-matrices here). Therefore the complete boundary
bootstrap
is closed in principle.

\subsec{Integral representations of various S-matrices.}

For comparison with results obtained from regularizations of the
sine-Gordon model it is useful to write integral representations of
the  boundary $S$-matrices
\smatr, \EQI\ and \EQII\ using the well-known formula
\eqn\main{\log\Gamma(z)=\int_0^{\infty}{dx\over x}e^{-x}
\left[z-1+{e^{-(z-1)x}-1\over 1-e^{-x}}\right], \qquad {\rm Re} z >
0. }
Suppose first that $1<2\xi/\pi<\lambda+1$ and denote
\eqn\Nn{n_{\ast}=\left[{\xi\over\pi}-{1\over 2}\right],}
where the square brackets mean the integer part of the number.
For such values of $\xi$
there are $n_{\ast}+1$ poles \poles\ in the physical strip, i.e. the
spectrum
of excitations contains boundary bound states. Correspondingly, there
is a
finite  number of $\Gamma$-functions in \smatr, \EQI, \EQII\ whose
arguments have negative real part so that formula \main\ is not
applicable.
Treating such  $\Gamma$-functions separately, we obtain the following
results:
\eqn\repI{
\eqalign{-i{d\over d\theta}\log \left[{P^+(\theta)\over R_0(\theta)}
\right]={2\lambda\over\pi}
\int_{-\infty}^{+\infty}&dx \cos\left({2\lambda\theta
x\over\pi}\right)
\cr
&\times\left[{\sinh(2\xi/\pi-2n_{\ast}-2)x\over\sinh x} +
{\sinh(\lambda-2\xi/\pi)x\over 2\sinh x \cosh \lambda x}\right],
\cr}}
\eqn\repII{\eqalign{-i{d\over d\theta}\log \left[ {P^+_{\beta_n}
(\theta)\over R_0(\theta)}\right] ={2\lambda\over\pi}
&\int_{-\infty}^{+\infty}dx \cos\left({2\lambda\theta
x\over\pi}\right)
\cr
&\times{\sinh(\lambda-2\xi/\pi)x - 2\cosh x
\sinh(\lambda+1+2n-2\xi/\pi)x
\over 2\sinh x \cosh \lambda x},  \cr}  }
\eqn\repIII{\eqalign{
-i{d\over d\theta}\log \left[{P^-_{\beta_n}(\theta)\over
R_0(\theta)}\right]&={2\lambda\over\pi}
\int_{-\infty}^{+\infty}dx \cos\left( {2\lambda\theta
x\over\pi}\right)
\left[ {\sinh(2n_{\ast}+2-2\xi/\pi)x\over\sinh x}  \right.\cr
&\left.+{\sinh(\lambda-2\xi/\pi)x - 2\cosh x
\sinh(\lambda+1+2n-2\xi/\pi)x
\over 2\sinh x \cosh \lambda x}\right] . \cr}  }
In the derivation of analogous representation for $P^-$ there are no
subtleties
because the "dangerous" $\Gamma$-functions cancel. We get
\eqn\repIIII{-i{d\over d\theta}\log \left[{P^-(\theta)\over
R_0(\theta)}
\right]={2\lambda\over\pi}
\int_{-\infty}^{+\infty}dx \cos\left({2\lambda\theta
x\over\pi}\right)
{\sinh(\lambda-2\xi/\pi)x\over 2\sinh x \cosh \lambda x}.  }
In the region $0<2\xi/\pi<1$
where there are no poles and no  boundary bound states in the
spectrum,
formula \repIIII\ is valid, too. The expression for $P^+$ can be
obtained
 from
\repI\ by setting formally $n_B\equiv n_{\ast}+1=0$, which gives
\eqn\repV{-i{d\over d\theta}\log \left[{P^+(\theta)\over R_0(\theta)}
\right]={2\lambda\over\pi}
\int_{-\infty}^{+\infty}dx \cos\left({2\lambda\theta
x\over\pi}\right)
{\sinh(\lambda+2\xi/\pi)x\over 2\sinh x \cosh \lambda x}. }
Note that if $2\xi/\pi>1$, the integral in \repV\ diverges.
Finally, we complete this list by the following two expressions:
\eqn\reepr{\eqalign{
&-i{d\over d\theta}\log \left[{P^{\pm}_{\delta_{N,n}}(\theta)\over
R_0(\theta)}\right]=-i{d\over d\theta}\log
\left[{P^{\pm}_{\beta_n}(\theta)\over
R_0(\theta)}\right]  + \cr
& +{2\lambda\over\pi}
\int_{-\infty}^{+\infty}dx \cos\left( {2\lambda\theta
x\over\pi}\right)\left[{\sinh({2\xi\over\pi}-2n_{\ast}-2)x\over\sinh x}
-{2\cosh x\sinh({2\xi\over\pi}+2N-\lambda-1)x\over 2\sinh x\cosh\lambda
x}\right].\cr }}
For the
integral representation
of $R_0$ see \FS.

\newsec{Exact solution of the regularized boundary sine-Gordon
model.}

\subsec{The XXZ chain with boundary magnetic field.}
The XXZ model in a boundary magnetic field

\eqn\XXZ{{\cal H}= {\pi-\gamma\over 2 \pi\sin\gamma}
\left[\sum_{i=1}^{L-1} \left(
\sigma^x_i \sigma^x_{i+1} + \sigma^y_i
\sigma^y_{i+1}+\Delta(\sigma^z_i
\sigma^z_{i+1}-1)\right) +h(\sigma_1^z-1)+h^{\p}(\sigma_L^z-1)
\right],}
was discussed in \ABBBQ, where its eigenstates were constructed using
the Bethe ansatz.  As usual, these  eigenstates ${\cal H}|n\>=E|n\>$
are linear combinations of
the states with $n$ down spins, located at
$x_1,...,x_n$ on the chain:
$$|n\>=\sum
f^{(n)}(x_1,...,x_n)|x_1,...,x_n\>.
$$

Consider for simplicity the case $n=1$. The wave-function  $f^{(1)}(x)$ reads
\ABBBQ:
\eqn\wf{\eqalign{f^{(1)}(x)&=[e^{-ik}+(h^{\p}-\Delta)]e^{-i(L-x)k} -
(k\to-k) =\cr
&=\left[{\sinh{1\over 2}(i\gamma+\alpha)\over
\sinh{1\over 2}(i\gamma-\alpha)}\right]^{L-x} {\sin\gamma\sinh{1\over
2}
(\alpha+i\gamma H^{\p})\over \sinh{1\over 2}(i\gamma-\alpha)
\sin{1\over 2}(\gamma+\gamma H^{\p})} -  (\alpha\to-\alpha),\cr}}
where we defined the new variables  as in
\ABBBQ: $\Delta=-\cos\gamma$, $k=f(\alpha, \gamma)$,
\eqn\Hh{\g H=f(i\g, -i\ln(h-\Delta))=-\g-i\ln{h-i\sin\g\over
h+i\sin\g}}
(and similarly for $H^{\p}$), and
\eqn\phase{f(a,b)=-i\ln\left[{\sinh{1\over 2}(ib-a)\over
\sinh{1\over 2}(ib+a)}\right].}
 When $h$ varies from  $0$ to $+\infty$, $\g H$ increases
monotonically
 from $-\pi-\g$ to $-\g$ according to \Hh\  if we take the
main branch
of the logarithm.

Denote $h_{th}=1-\cos\g$. This
``threshold'' value of $h$ corresponds to $\g H=-\pi$;  its meaning
will become clear below. When $h$ varies from $-\infty$ to $0$, $\g
H$
increases monotonically  from $-\g$ to $\pi-\g$. For the purposes of
the
present work we confine our attention to the region $h, h^{\p}>0$ and
choose
$\g\in(0,{\pi\over 2})$. Other regions in the parameter space
can be obtained using the discrete symmetries of the Hamiltonian
\XXZ:
$\sigma^z\to -\sigma^z$ on each site or on the odd sites only.
The parameter $k$ in \wf\ is not arbitrary, but satisfies the Bethe
equation  \ABBBQ:
\eqn\BAEV{
e^{i(2L-2)k}{(e^{ik}+h-\Delta)(e^{ik}+h^{\p}-\Delta)\over(e^{-ik}+h-
\Delta)(e^{-ik}+h^{\p}-\Delta)}=1,}
or
\eqn\BAEI{\left[{\sinh{1\over 2}(\alpha-i\gamma)\over
\sinh{1\over 2}(\alpha+i\gamma)}\right]^{2L}
{\sinh{1\over 2}(\alpha-i\gamma H)\sinh{1\over 2}(\alpha-i\gamma
H^{\p})
\over
\sinh{1\over 2}(\alpha+i\gamma H)\sinh{1\over 2}(\alpha+i\gamma
H^{\p})}=1.}
Note that the wave-function  \wf\ depends on $H$ implicitely through
the solution of the Bethe equation \BAEI\
$\alpha(H,H^{\p})$. Besides, one can multiply the amplitude \wf\ by
any
overall scalar factor depending on $\alpha$, $L$, $H$ and $H^{\p}$.
The Bethe equations in the sector of arbitrary $n>1$ can be found in \ABBBQ.

\subsec{The Bethe equations for the inhomogeneous XXZ chain.}

The real object of interest for us is actually the  inhomogeneous
six-vertex model with boundary magnetic field
on an open strip. The inhomogeneous six-vertex model is obtained by
giving an alternating
imaginary part $\pm i\Lambda$  to the spectral parameter on
alternating vertices of the six-vertex model
\refs{\RS, \DDV}.  It was argued in \FS, generalizing known
results for the periodic case \DDV\  that this  model
on an open strip provides in the scaling limit  $\Lambda,
L\to\infty$, lattice spacing $\to 0$
a lattice regularization of \SG,
with $\beta^2=8\g$ and  a value of $\varphi_0$ at the boundary
related to
the magnetic field. The reader can find more details on the model in
the
references; it is actually closely related to the XXZ chain we
discussed above. In particular,
the wave function can be
expressed in
terms of the  roots  $\alpha_j$
 of the Bethe equations  \refs{\ABBBQ,\RS}:
\eqn\origbe{\eqalign{
&\left[{\sinh{1\over 2}(\alpha_j+\Lambda-i\g)\over
\sinh{1\over 2}(\alpha_j+\Lambda+i\g)}
{\sinh{1\over 2}(\alpha_j-\Lambda-i\g)\over
\sinh{1\over 2}(\alpha_j-\Lambda+i\g)}\right]^L
{\sinh{1\over 2}(\alpha_j-i\g H)\over\sinh{1\over 2}(\alpha_j+i\g
H)}\times\cr
& {\sinh{1\over 2}(\alpha_j-i\g H^{\p})\over
\sinh{1\over 2}(\alpha_j+i\g H^{\p})} =\prod_{ m\neq j}
 {\sinh{1\over 2}(\alpha_j-\alpha_m-2i\g)\over
\sinh{1\over 2}(\alpha_j-\alpha_m+2i\g)}
 {\sinh{1\over 2}(\alpha_j+\alpha_m-2i\g)\over
\sinh{1\over 2}(\alpha_j+\alpha_m+2i\g)}.    \cr} }
By construction of the Bethe-ansatz wave function, $\alpha_j>0$.
Note that the solutions of \origbe\ $\alpha_j=0, i\pi$ should be
excluded
  because the wave function
vanishes identically in this case.  The analysis of solutions of the
\origbe\
is very similar to the case of the XXZ chain in a boundary magnetic
field.  We consider the regime $0<\g<\pi/2$, which falls
into the
attractive regime $0<\beta^2<4\pi$ in the sine-Gordon model \SG.
We set hereafter $\g=\pi/t$ and for technical simplicity restrict
$t$
to be positive integer. In the limit $L\to\infty$ this  constraint
implies that
in the bulk
only the strings of length from 1 to $t-1$ are allowed, together with the
anti-strings.

Taking the logarithm  of eq. \origbe, one obtains:
\eqn\foral{\eqalign{L\left[f(\alpha_j+\Lambda,\gamma)+
f(\alpha_j-\Lambda,\gamma)\right]+
f(\alpha_j,\gamma H)+f(\alpha_j,\gamma H^{\p})\cr
=2\pi l_j + \sum_{ m\ne
j} \left[f(\alpha_j-\alpha_m,2\gamma) +
f(\alpha_j+\alpha_m,2\gamma)\right],\cr}}
where $l_j$ is an integer.
We also recall the formula for the
eigenenergy associated with the roots $\alpha_j$ \refs{\ABBBQ, \RS}
\eqn\Energy{E={2(\pi-\gamma)\over\pi}\sum_{\alpha_j}[f^{\p}
(\alpha_j+\Lambda, \gamma) + f^{\p}(\alpha_j-\Lambda, \gamma)].}

\subsec{Thirring model with boundary.}

Like the bulk sine-Gordon model is a bosonized version
of the bulk massive Thirring model, one can expect that the boundary
sine-Gordon model is a bosonized version of the
Thirring model with  certain boundary conditions.
The quickest way to identify this boundary
Thirring model is to use the Bethe ansatz equations \origbe. Write
the most general $U(1)$-invariant boundary interaction
\eqn\Thir{\eqalign{H_{T}=
\int_0^L dx & [-i\psi_1^+\psi_{1x} + i\psi_2^+\psi_{2x} +
m_0\psi_1^+\psi_2+m_0\psi_2^+\psi_1+2g_0\psi_1^+\psi_2^+\psi_2\psi_1]
\cr
&+\sum_{ij}a_{ij}\psi_i^+\psi_j(0)+\sum_{ij}a^{\p}_{ij}\psi_i^+
\psi_j(L).\cr}}
The entries of the $2\times 2$ matrices $A=\{a_{ij}\},
A^{\p}=\{a^{\p}_{ij}\}$
can be determined up to one arbitrary parameter $\varphi$ by the
hermicity
of $H_{T}$ and the consistency of the boundary conditions $({\rm det}
A=0)$.
For the left boundary, the matrix $A$ looks like
\eqn\lmatr{A={1\over 2\sin\phi}\pmatrix{e^{-i\phi}& 1 \cr
                        1& e^{i\phi}\cr}}
and the boundary condition reads $\psi_1(0)=-e^{i\phi}\psi_2(0)$
(similarly for the right boundary).

To find the eigenvectors of the Hamiltonian \Thir, $H_{T}\Psi=E\Psi$,
one can use the same wave-functions as for the bulk Thirring model
\Kor, and modify them by analogy with the example of XXZ chain in a
boundary
magnetic field \ABBBQ. This way one gets the equations
for the set of rapidities
$\alpha_j$ :
\eqn\origbeI{\eqalign{ e^{2im_0L\sinh\alpha_j}=&
{\cosh{1\over 2}(\alpha_j+i\phi)\over\cosh{1\over
2}(\alpha_j-i\phi)}
 {\cosh{1\over 2}(\alpha_j+i\phi^{\p})\over
\cosh{1\over 2}(\alpha_j-i\phi^{\p})}\cr & \times\prod_{ m\neq
j}
 {\sinh{1\over 2}(\alpha_j-\alpha_m-2i\g)\over
\sinh{1\over 2}(\alpha_j-\alpha_m+2i\g)}
 {\sinh{1\over 2}(\alpha_j+\alpha_m-2i\g)\over
\sinh{1\over 2}(\alpha_j+\alpha_m+2i\g)}, \cr}    }
where $\gamma$ is related to $g_0$ in the usual way \Kor. These
equations look quite similar to \origbe. The mapping can be made
complete by taking in \origbe\ the limit $\Lambda\rightarrow\infty$
with the identification $m_0=4e^{-\Lambda}\sin\gamma$.

The
derivation of
these equations is rather cumbersome, therefore to illustrate the
procedure
we comment on the simplest case of one-particle sector, which is
nevertheless
sufficient to obtain the form of the boundary terms in \origbeI.
We make an ansatz $\Psi=\int_0^L dy
\chi^{\lambda}(y)\psi^+_{\lambda}(y)|0\>$,
where $\lambda$ is the spinor index, $\chi(y)$ is the wave-function
and $|0\>$ is the unphysical vacuum annihilated by $\psi_{\lambda}$.

The equation $H_{T}\Psi=E\Psi$ reduces to ($\sigma_i$ are the Pauli
matrices):
\eqn\IwI{ -i\sigma_3{\partial\over \partial x}\vec\chi
+m_0\sigma_1\vec\chi
+A\vec\chi\delta(x)+A^{\p}\vec\chi\delta(x-L)=E\vec\chi.}
We look for the solution of \IwI\ in the form
\eqn\IIwI{\pmatrix{\chi_1\cr\chi_2\cr}=a(\alpha)\pmatrix{e^{-\alpha/2}
\cr
e^{\alpha/2}\cr}e^{im_0x\sinh\alpha} -
a(-\alpha)\pmatrix{e^{\alpha/2}\cr
e^{-\alpha/2}\cr}e^{-im_0x\sinh\alpha}. }
Substituting it into \IwI\ we get $E=m_0\cosh\alpha$ and,
besides, two boundary conditions
to be solved. The first one, at $x=0$, determines the form of the
 factor $a(\alpha)=\cosh{1\over 2}(\alpha-i\phi)$, while the
second one
at $x=L$ gives rise to the Bethe equation
$$e^{2im_0L\sinh\alpha}=
{\cosh{1\over 2}(\alpha+i\phi)\over\cosh{1\over
2}(\alpha-i\phi)}
{\cosh{1\over 2}(\alpha+i\phi^{\p})\over
\cosh{1\over 2}(\alpha-i\phi^{\p})},$$
which determines $\alpha$. Comparing the Bethe equation \origbe\ with
\origbeI\ and using the relation
between $\xi$ and $H$ obtained below in section 5
 we find the relation between the boundary
parameters $\phi$ and $\varphi_0$ in the Hamiltonians \Thir\ and
 \SG\ respectively:
$$\phi=\beta\varphi_0-\beta^2/8.$$
Thus, the integrable boundary condition
for the $U(1)$-invariant boundary Thirring model reads:
\eqn\intconI{\psi_2(0)=-e^{i\beta^2/8-i\beta\varphi_0}\psi_1(0).}

It would be interesting to obtain the result \intconI\ directly
from the Hamiltonian \SG\ using
an extension of the Coleman-Mandelstam bosonization technique to
the case with boundary. However, to our knowledge such an extension
has not been developed yet. The naive application of the known
Coleman-Mandelstam ``bulk'' formulas doesn't give the factor
$e^{i\beta^2/8}$ in \intconI, which seems to be some kind of
``boundary anomaly''.

\newsec{Solutions of the Bethe ansatz equations with boundary terms.}

As is well known in the case of the bulk Thirring model or
equivalently
the periodic XXZ chain, the  bound states are associated with
various types of solutions of the Bethe ansatz equations involving
in general complex roots \Kor. By analogy, we expect the
boundary bound states  to
correspond to new solutions
made possible by the boundary terms.

Consider first the example of the XXZ chain
as given in section 3.1. Since our goal is to study
purely boundary effects, we will look for  the solutions of the Bethe equations
that give rise to
a wave-function localized at $x=0$ or $x=L$ and exponentially
decreasing
away from the boundary.

The states described by such wave-functions will be referred to
as the ``boundary bound states'' below.  For this, one should have
$\alpha$ purely imaginary in \wf.
We consider here the limit of $L$
large, when the left and the right boundaries can be treated
independently
and the overlap of the corresponding wave-functions is negligibly
small
(for the physical applications it is necessary to take the scaling
limit
anyway).
 In the limit $L\to\infty$ it is easy to check that there are two such
solutions
to \BAEI: $\alpha=i\alpha_0=-i\gamma H +
i\varepsilon(L,H,H^{\p}) $ and
$\alpha=i\alpha_0^{\p}=-i\gamma
H^{\p}+i\varepsilon^{\p}(L,H,H^{\p})$,
where $\varepsilon\sim
\exp(-2\kappa L)$ and we defined $\kappa>0$ as
$$ e^{-\kappa}=\left|{\sin{1\over 2}(-\gamma H-\gamma)\over
\sin{1\over 2}(-\gamma H+\gamma)}\right|$$
(similar relations are assumed for $\varepsilon^{\p}, \kappa^{\p}$).  Solution
$\alpha_0^{\p}$ gives a wave-function \wf\ localized at
$x=L$: $f^{(1)}(x)\sim  e^{-\kappa^{\p}(L-x)}$. Solution
$\alpha_0$ gives a wave-function localized at $x=0$,
$f^{(1)}(x)\sim  e^{-\kappa x}$, provided we renormalize  the
wave-function
\wf:
\eqn\wfI{f^{(1)}\rightarrow f^{(1)}\left[\sinh{1\over
2}(\alpha-i\gamma H)
\sinh{1\over 2}(\alpha+i\gamma H)\right]^{1/2}.}
In the special case $H=H^{\p}$ there is  only one proper solution
$\alpha=i\alpha_0=-i\g H +i\varepsilon(L,H)$ with
$\varepsilon\sim\exp(-\kappa L)$.
The  wave-function  \wf\ behaves as the superposition
of the "left" and the "right" boundary bound states,
$f^{(1)}\sim(e^{-\kappa x} + e^{-\kappa(L-x)})$.
Note that the boundary bound state appears in the above example
only when the boundary magnetic field is large enough: namely,
$h>h_{th}$\foot{More generally, the criterion of existence of boundary bound
state
solutions
allows us to determine  threshold fields for any
$\Delta$.
For this, let us examine \BAEV. The parameter $k$ is defined modulo
$2\pi$,
therefore we restrict it to lie withing $k\in (0,2\pi)$. Two
possibilities
$k=ia$ and $k=\pi+ia$, where $a>0$, lead to two different threshold
fields,
determined by the fact that the denominator in \BAEV\ should vanish:
$$h_{th}^{(1)}=\Delta +1, \qquad h_{th}^{(2)}=\Delta-1,$$
 and the regions with boundary bound states are $h<\Delta-1, \quad
h>\Delta+1$.
When $\Delta>1$, there are two different threshold fields, in
agreement with the results of Jimbo et al \miwa. In the region
$|\Delta|<1$,
discussed here, there is only one threshold field $h_{th}^{(1)}$.}
. This follows from the fact that $\alpha$ should be such
that
$0<\alpha_0<\pi$.

Now, consider the equations for the inhomogeneous model \origbe.
The basic boundary 1-string solution to \origbe\ is
$\alpha=i\alpha_0=-i\g H+i\epsilon$, provided that $0<\alpha_0<\pi$.
This solution is possible due to an argument very similar
to the one used in the bulk: as $L\to\infty$,  the two first terms
of \origbe\ decrease exponentially fast, while the third increases
exponentially fast, and $\epsilon\sim \exp(-2\kappa L)$ with
\eqn\Defkap{e^{-\kappa}={ \sinh^2{\Lambda\over 2}+
\sin^2{\alpha_0-\g\over 2}
\over \sinh^2{\Lambda\over 2}+ \sin^2{\alpha_0+\g\over 2}}.}
Recall that for the  bulk problem when there is no boundary term, the right
hand side of \origbe\ would have to decrease exponentially, forcing the
existence of a ``partner''  root at $\alpha-2i\gamma$.

One can construct similarly  boundary n-strings which  consist of
the points $i\alpha_0, i\alpha_0+2i\g, ... , i\alpha_0+2i(n-1)\g$
(see  figure 1). By convention $n=0$ means there is no boundary string, that is
all complex solutions are in the usual bulk strings.
The possible values of $n$ are restricted  by the fact that the upper
point of the complex should be below $i\pi$:
$\max(n)=[{\pi-\alpha_0\over
2\g}]+1$, where the square bracket denotes the integer part.  To show
that
the boundary n-string is indeed a solution to \origbe, we introduce
infinitesimal  corrections $\varepsilon_i$ to the positions of the
points of
complex \Gaud. Taking the modulus of both sides of \origbe\ with
$\alpha_j=
i\alpha_0+2ik\g$ and multiplying equations for $k=0,...,n-1$
we obtain $\exp\{-2L(\kappa_1+\kappa_2+...+\kappa_n)\}\sim
\varepsilon_1$, where $\varepsilon_1$ denotes the correction to the
point
$i\alpha_0$. The behavior of the remaining $\varepsilon_k$ follows
from
$\varepsilon_1$ by recursion. For example, for the 2-string
$\varepsilon_2$ is given by  $|\varepsilon_1
-\varepsilon_2|\sim \exp(-2L\kappa_2)$
\foot{
Note that associated  with  each
boundary n-string there
is also the solution to \origbe\ obtained by complex
conjugation of  all $\alpha'$s.  The existence of such a "mirror
image" is the consequence of the symmetry of equations \origbe\
and it is of no importance to physics. In the bulk case, it is easy to show
\BdVV\ that all solutions are invariant under complex conjugation, but this
result does not hold here.
In fact, a solution which  both the boundary n-string and its
mirror image would lead to a vanishing wave-function.}.

Additional boundary strings can be obtained by adding the roots
$i\alpha_s$
below $i\alpha_0$ so that $i\alpha_s=i\alpha_0-2is\g$, with
$s=1,2,...,N$
(see  figure 2).
Together with the boundary n-string above $\alpha_0$, they form the
complex
which we call boundary $(n,N)$-string. To analyze the existence of
such
complexes  as the solutions of \origbe\ we introduce  as before the
infinitesimal
corrections $\varepsilon_s$ to the roots $\alpha_s$, where now
 $s=n, n-1,..., 1,
-1, -2,..., -N$. Then, the equations \origbe\ with
$\alpha_j=i\alpha_s$ tell us
that the range of $N$ should be
\eqn\RangeN{ {\alpha_0\over 2\g}<N<{\pi+\alpha_0\over 2\g}.}
In other words, the inequality \RangeN\ states that the lowest root
of the
boundary string should be below the axis ${\rm Im}\alpha=0$ and above
the axis ${\rm Im}\alpha=-i\pi$.
Another constraint  follows if we multiply the equations \origbe\ for
all the roots of boundary $(n,N)$-string. This gives
$\exp(-2L\sum\kappa_s)
=\varepsilon_1$. So, one should have $\sum\kappa_s>0$. The latter sum
can
be easily evaluated if one uses the expression \Defkap\ simplified in
the
limit $\Lambda\to\infty$: $\kappa=4e^{-\Lambda}\sin\g\sin\alpha_0$.
The constraint obtained in such a way forces the number of roots
above ${\rm Im}\alpha=0$ axis in the boundary string to be greater
than the
number of roots below ${\rm Im}\alpha=0$.

We have not been able to find any reasonable additional solution to
the
Bethe ansatz equations.
The  two sets of
boundary strings we have encountered  appear
to be in one to one correspondence with the boundary bound states
identified in section 2 using the bootstrap approach. To clarify
this identification we now compute related  masses and S-matrices.

\newsec{$S$ matrices and bound state properties from the exact
solution.}

\subsec{Bare and physical Bethe ansatz equations.}

The "bare" Bethe equations follow from taking the derivative of
$\foral$.
Defining $2L(\rho_k+\rho_k^h)d\alpha$ to be the number of roots
in the interval $d\alpha$, one obtains coupled integral equations
for the densities of strings $\rho_1,...,\rho_{t-1}$ and anti-strings
$\rho_a$:
\eqn\bba{\eqalign{2\pi (\rho_k + \rho_k^h)&=  {1\over 2}
p_k^{\p} - f^{\p}_{ka}\ast\rho_a -
\sum_{l=1}^{t-1}f^{\p}_{kl}\ast\rho_l
+{1\over 2L}(u_k-\omega f^{\p (L)}_{n;k}-
\omega^{\p}f^{\p (R)}_{n^{\p};k})\cr
2\pi(\rho_{a} +\rho_a^h)&=-{1\over 2}p_a^{\p}+f^{\p}\ast\rho_a
+\sum_{l=1}^{t-1}f^{\p}_{al}\ast\rho_l+
{1\over 2L}(u_a+\omega f^{\p (L)}_n + \omega^{\p} f^{\p
(R)}_{n^{\p}})\cr}}
where $\ast$ denotes convolution:
$$ f\ast g(\alpha)= \int_{-\infty}^{\infty} d\beta f(\alpha-\beta)
g(\beta).$$
These densities are originally defined for $\alpha>0$, but the
equations allow
us to define $\rho_k(-\alpha)\equiv\rho_k(\alpha)$ in order to
rewrite the
integrals to go from $-\infty$ to $\infty$.  If we totally neglect
the boundary
terms (terms $\sim L^{-1}$) in  \bba, we will end up with the same
equations
as for the periodic inhomogeneous six-vertex model \RS.  The various
kernels
and sources in \bba\ are defined as follows:
$$p_a(\alpha)=f(i\pi+\alpha+\Lambda, \g)+f(i\pi+\alpha-\Lambda,
\g),$$
$$p_k(\alpha)=\sum_{\alpha_i}f(\alpha_i+\Lambda, \g)+
f(\alpha_i-\Lambda, \g),$$
where the sum in the last expression is taken over the rapidities of
the
bulk k-string root centered on $\alpha$.

The kernels $f_{kl}$
are the phase shifts of bulk k-string on bulk $l$-string obtained by
summing
\phase\
over the rapidities of string roots. The boundary terms are:
$$u_a(\alpha)=-2f^{\p}(2\alpha, 2\g)-f^{\p}(\alpha+i\pi, \g H)
-f^{\p}(\alpha+i\pi, \g H^{\p})-2\pi\delta(\alpha),$$
$$u_k(\alpha)=\sum_{\alpha_i}[2f^{\p}(2\alpha_i, 2\g)+
f^{\p}(\alpha_i, \g H)+f^{\p}(\alpha_i, \g
H^{\p})]-2\pi\delta(\alpha),$$
(the sum above is over the roots of bulk k-string centered on $\alpha$),
$$f_n^{(L,R)}(\alpha)=\sum_{\alpha_i}
f(i\pi+\alpha-\alpha_i, 2\g)
+f(i\pi+\alpha+\alpha_i, 2\g),$$ and $\alpha_i$
denotes the rapidities of the roots in the boundary n-string.
$$f^{(L,R)}_{n;k}(\alpha)=\sum_{\alpha_i}\sum_{\alpha_j}
f(\alpha_j-\alpha_i, 2\g)
+f(\alpha_j+\alpha_i, 2\g),$$
where $\alpha_i$ denotes the roots in the boundary n-string, while
$\alpha_j$ denotes the roots in the bulk k-string centered on $\alpha$.
The parameters
$\omega, \omega^{\p}$ are equal to $1$ or $ 0$, depending on whether
the boundary string is present or not .
In our case, $0<\g<\pi/2$, the ground state
of the periodic inhomogeneous XXZ chain is filled with anti-strings.
The physical Bethe equations are obtained
\ref\DL{C. Destri, J. H. Lowenstein, Nucl. Phys. B205 (1982) 369.},\ref\KR{A.
N. Kirillov, N. Reshetikhin, J. Phys. A20 (1987) 1565, 1587.} by
eliminating the "non-physical" density $\rho_a$ from
 the right-hand side of \bba. This is done simply
by solving  for
$\rho_a$ in the last equation in \bba\ and substituting it into
the others. The result is
\eqn\physbe{\eqalign{
2\pi(\rho_k+\rho_k^h)&={1\over 2}p^{\p}_k
+{1\over 2}{f^{\p}_{ak}\over 2\pi-f^{\p}}\ast p^{\p}_a +\cr
&+ {f^{\p}_{ak}\over 2\pi-f^{\p}}\ast 2\pi\rho_a^h - \sum_{l=1}^{t-1}
\left(f^{\p}_{kl}+{f^{\p}_{ak}f^{\p}_{al}\over
2\pi-f^{\p}}\right)\ast\rho_l
+{1\over 2L}U_{n,n^{\p}; k}, \cr
2\pi(\rho_a+\rho_a^h)&=-{1\over 2}{2\pi p^{\p}_a\over 2\pi-f^{\p}}
-{f^{\p}\over 2\pi-f^{\p}}\ast 2\pi\rho_a^h + \sum_{l=1}^{t-1}
{f^{\p}_{al}\over 2\pi-f^{\p}}\ast 2\pi\rho_l + {1\over
2L}U_{n,n^{\p}; a},
\cr }}
where
\eqn\defI{U_{n,n^{\p}; a}=2\pi{
u_a+\omega f^{\p (L)}_n + \omega^{\p} f^{\p (R)}_{n^{\p}}
\over 2\pi-f^{\p}}, }
\eqn\defII{U_{n,n^{\p}; k}=u_k-\omega f^{\p (L)}_{n;k}-
\omega^{\p}f^{\p (R)}_{n^{\p};k}-f^{\p}_{ak}\ast U_{n,n^{\p};
a}/2\pi,}
and different products  (ratios) of kernels are defined through their
Fourier transforms.

\subsec{The mass spectrum of boundary bound states.}

We assume at first
that the ground state is built by filling up the Dirac sea
with anti-strings, as in the case of the periodic XXZ chain.
We will see below that this is not always true.
The presence of the boundary strings in the Bethe equations
deforms the distribution of  roots and modifies
the density of the Dirac sea $\rho_a$ by a term $\delta\rho_a/2L$
of  order $L^{-1}$. With the boundary n-string, the Bethe equation
for the density of the Dirac sea particles $\tilde{\rho}_a$ is
\eqn\beI{{1\over 2} p_a^{\p}(\alpha)-{1\over
2L}u_a(\alpha)=\int_{-\infty}^{
+\infty}f^{\p}(\alpha-\beta)\tilde{\rho}_a(\beta)d\beta
-2\pi\tilde{\rho_a}
(\alpha) + {1\over 2L}f_n^{\p}(\alpha),}
where $f_n$ was defined above.
Subtracting from \beI\ the equation for the density of the Dirac sea
alone,
\eqn\beII{{1\over 2} p_a^{\p}(\alpha)-{1\over
2L}u_a(\alpha)=\int_{-\infty}^{
+\infty}f^{\p}(\alpha-\beta)\rho_a(\beta)d\beta -2\pi\rho_a(\alpha)
,}
one obtains the equation for $\delta\rho_a$:
\eqn\beIII{0=-\int_{-\infty}^{
+\infty}f^{\p}(\alpha-\beta)\delta\rho_a(\beta)d\beta +
2\pi\delta\rho_a(\alpha) - f_n^{\p}(\alpha).}
$$\delta\rho_a\equiv 2L(\tilde{\rho}_a-\rho_a).$$
The solution to \beIII\ can be written in terms of the Fourier
transform
$\delta\hat{\rho}_a(k)=\int d\alpha e^{ik\alpha}\delta\rho_a(\alpha)$
as follows:
\eqn\delro{\delta\hat{\rho}_a(k)={\hat{f}_n^{\p}(k)\over
 2\pi-\hat{f}^{\p}(k)}.}
For the boundary n-string  $i\alpha_0+2i\gamma s$,
$s=0,1,...,n-1$ we obtain
\eqn\ftI{\hat{f}_n^{\p}(k)=-2\pi{4\cosh \gamma k \sinh n\gamma k
\cosh(\alpha_0 +\gamma n -\gamma)k \over \sinh \pi k}, }
\eqn\deformI{\delta\hat{\rho}_a(k)=-{2\cosh \gamma k \sinh n\gamma k
\cosh(\alpha_0
+\gamma n -\gamma)k \over \sinh \gamma k \cosh(\pi-\gamma)k}, }
where we used
$$\hat{f}^{\p}(k)=2\pi{\sinh(\pi-2\gamma)k\over\sinh \pi k}.$$
Expressions \ftI, \deformI\ are valid for the n-strings with
$n=1,2,...,
\left[{t+H\over 2}\right]$. For the longest n-string with
$n=\left[{t+H\over 2}\right]+1\equiv n_{\ast}+1$  the Fourier
transforms
$\hat{f}_n^{\p}, \delta\hat{\rho}_a$ differ from \ftI, \deformI:
\eqn\ftII{\eqalign{
\hat{f}_{\ast}^{\p}(k)=&2\pi{2\sinh(\pi-2\gamma)k \cosh
(\alpha_0+2\gamma n_{\ast}-\pi)k \over \sinh \pi k} \cr
& - 2\pi{4\cosh \gamma k \sinh n_{\ast}\gamma k \cosh(\alpha_0
+\gamma n_{\ast}-\gamma)k\over \sinh \pi k}, \cr} }
\eqn\deformII{\eqalign{
\delta\hat{\rho}_a(k)=&{\sinh(\pi-2\gamma)k \cosh
(\alpha_0+2\gamma n_{\ast}-\pi)k \over \sinh \gamma k \cosh
(\pi-\gamma)k} \cr
& - {2\cosh \gamma k \sinh n_{\ast}\gamma k \cosh(\alpha_0
+\gamma n_{\ast}-\gamma)k\over  \sinh \gamma k \cosh
(\pi-\gamma)k}.  \cr} }
The conserved $U(1)$ charge in the boundary XXZ chain is  the
total projection of the spin on the z-axis. In the thermodynamic limit
the charge of the boundary n-string with respect to the vacuum
is determined by \ABBBQ:
\eqn\charge{ Q_n=n+\int_0^{+\infty}2L\tilde{\rho}_ad\alpha
- \int_0^{+\infty}2L\rho_ad\alpha=n+{1\over
2}\int_{-\infty}^{+\infty}
\delta\rho_a d\alpha=n+{1\over 2}\delta\hat{\rho}_a(0).}
Using \deformI, we obtain for the n-string  $ Q_n=0$, and for
the longest boundary string Eq. \deformII\ yields $Q_{\ast}=\pi/
2\gamma$.
Similarly, the mass of the boundary strings
in the thermodynamic limit according to \Energy\ is given by
\eqn\ener{m_n=h_n+\int_0^{+\infty}2L\tilde{\rho}_a
h_ad\alpha - \int_0^{+\infty}2L\rho_ah_ad\alpha=h_n+{1\over 2}
\int_{-\infty}^{+\infty}h_a\delta\rho_a d\alpha,}
where the expression for $h_a$ is
$$ \hat{h}_a(k)={2(\pi-\g)\over\pi}\hat{p}_a^{\p}=
-4(\pi-\g){2\sinh \gamma k \cos
\Lambda k
\over \sinh \pi k}$$
and the soliton mass \RS\
$$ m=4e^{-{\Lambda\pi\over 2(\pi-\gamma)}}.$$

We obtain in the limit $\Lambda\to\infty$
\eqn\enerI{m_n=m\left[\sin{\pi\over 2\lambda}\left(2n-1-H\right)+
\sin{\pi\over 2\lambda}\left(H+1\right)\right],}
\eqn\enerII{m_{\ast}=m\sin{\pi\over 2\lambda}\left(H+1\right).}
Since the parameter $H$ varies in the interval $-\lambda-1<H<-1$,
the mass of
the longest string $m_{\ast}$ \enerII\ is always negative, while
the other boundary strings have positive masses \enerI. This means
that
the vacuum we have been working with is an unstable one in the region
$-t<H<-1$ $(h>h_{th})$.  To cure the
situation we define a new correct  ground state by attributing the
longest
boundary string  to the Dirac sea.  The boundary excitations are
obtained by
succeessive removing  of particles from the top of the longest
boundary string.
The charge and mass of such excitations with respect to the correct
ground
state
are given by
\eqn\correct{Q_n=-{\pi\over 2\gamma}, \qquad m_n=
m\cos{\pi\over 2\lambda}(\lambda+1+H-2n), \qquad n=0,1,...,n_{\ast}.}
Note that the number of excitations \correct\ is equal to the
number of particles in
the longest boundary string, $n_{\ast}+1$. The charge of such
boundary
excitations is equal to the charge of the hole in the Dirac sea. We
identify
a hole with a sine-Gordon soliton, and the boundary excitations
described above, with the boundary bound states  $|\beta_n\> $ \poles. Their
masses and
charge \correct\  and the counting coincide provided that
\eqn\param{t+H+1={2\xi\over\pi},}
and the lattice charge $Q$ is properly
normalized. This expression is in fact valid for all values of $h>0$.
The authors of \FS, deriving this relation in the region $h<h_{th}$,
obtained  a different expression because they
 used different branch of logarithm in \Hh.

In the above discussion we considered the boundary bound states
related
to  one of the boundaries (say, the left one). In principle, one
should include
into the ground state the longest boundary string
$i\alpha^{\p}_0+2i\g l,
\quad l=0,1,...,\left[{t+H^{\p}\over 2}\right]$, corresponding to the
right
boundary as well. The energy of the excitations
due to both boundary strings is a superposition of energies of the
form
\correct. When $H=H^{\p}$, these two  boundary strings overlap and
the
usual Bethe wave-function vanishes. However on physical grounds we do not
expect anything special to happen when the boundaries are identical.
So, in such a case
 one should use as a wave function a properly renormalized
version of the
limit $H\to H^{\p}$
of the
usual Bethe wave function.

When the magnetic field varies, the above picture indicates
a   qualitative change in the
structure
of the  ground state at values  $H=-t, -t+2, -t+4,...$. At these values,
the mass of the bound state with the highest mass approaches  the soliton mass
and it becomes unstable.
As discussed  in \GZ\ for the Ising case,
this decay corresponds to large
  boundary fluctuations that propagate deeply into the bulk.

 The mass of the boundary
$(n,N)$-string with respect to the correct vacuum  can be calculated
analogously.
The result is:
\eqn\enerNn{m_{n,N}=m\cos({\xi\over\lambda}-{\pi\over 2\lambda})
+m\cos({\xi\over\lambda}-{2n+1\over 2\lambda}\pi)
-m\cos({\xi\over\lambda}+{2N-1\over 2\lambda}\pi),}
where we used \param\ to express $H$ interms of $\xi$.
This result is rather confusing to us, because the above
mass does not correspond in general to one of the bound state masses
found in the bootstrap apporach. It can be considered as a sum of such masses,
hinting that the $(n,N)$ string describes actually coexisting bound states,
but the corresponding boundary S-matrix does not allow such an interpretation.
We  are forced  (but see the conclusion) to consider that only the
$(n,N)$-strings with $n=n_{\ast}+1$
occur, that is
the physical excitations  are built by adding
roots to the ground state configuration below $i\alpha_0$.
The charge and energy of such excitations
with respect to the correct vacuum is given by
\eqn\enerNnI{Q_N={\pi\over 2\g}-{\pi\over 2\g}=0, \qquad
m_{N}=m\cos({\xi\over\lambda}-{\pi\over 2\lambda})
-m\cos({\xi\over\lambda}+{2N-1\over 2\lambda}\pi),}
These coincide with   the charge and  mass
of the boundary bound state $|\delta_{n=0,N}\>$
\massnew. The
range
of $N$ \RangeN\ agrees with the range of corresponding parameter in
\newpoles.

\subsec{Boundary $S$ matrices.}

It remains to check that the boundary S-matrices obtained above by
the bootstrap approach  coincide with  those of the lattice
model. To extract the boundary
S-matrices from the Bethe equations we will follow the discussion of
\FS.  Briefly, the idea of the method is the following.  The
physical
excitations of the  lattice model in the limit $\Lambda\to\infty$
can be thought
of as  relativistic quasi-particles with  rapidities $\t_i$. The
integrability
implies
that  the set $\{\t_i\}$ is conserved. Moreover, if the scattering
matrices are
diagonal, each  particle preserves its rapidity. Assuming that this
is the
 case, the quantization of a gas of ${\cal N}$ quasi-particles on
an interval of
lenght $L$ results in  the  integral equations for the set of allowed
rapidities
\FS:
\eqn\sba{2\pi(\rho_b+\rho_b^h)= m_b \cosh\t + \sum_{c=1}^p
\varphi_{bc}*\rho_c+ {1\over 2L}
\Theta_b,}
where subscript stands for the type of particle, and
\eqn\forphi{\eqalign{\varphi_{bc}(\t)&=-i {d\over d\t}
\ln S_{bc}(\t)\cr
\Theta_{b}(\t)&=-i {d\over d\t} \ln R_{\beta}^{b(L)}(\t)
-i {d\over d\t} \ln R_{\beta^{\p}}^{b(R)}(\t)
+ i {d\over d\t} \ln S_{bb}(2\t)
-2\pi \delta(\t).\cr}}
Equations \sba\ should be compared with the physical BE \physbe,
which gives
bulk and boundary S-matrices.  We will confine our attention to the
boundary
S-matrices only, keeping track of those terms  in \defI, \defII,
\forphi,
which depend on the boundary magnetic field (the field-independent
terms
contribute to $R_0$ and their agreement has been shown in \FS).
The discussion for the left boundary is completely parallel to that
of the
right
one. Also, it is
sufficient to consider only $b=$soliton  and $b=$anti-soliton in
\sba. We
identify a hole in the anti-string distribution in \physbe\ with a
soliton
in \sba, and $(t-1)$-string with  an anti-soliton. Below we give
explicit expressions only  for
the
 kernels in \physbe\ which we need
for our
analysis. The other expressions are listed in \FS.
Suppose first that $h<h_{cr}$ ($-t-1<H<-t$). This corresponds to the
case
without boundary excitations in the spectrum, $\xi<\pi/2$. Choose
$\omega=\omega^{\p}=0$ in \physbe. Then
$$\hat{u}_a^{(L)}(k)=2\pi{\sinh(2\pi+\g H)k\over \sinh\pi k}+\ldots$$
(we omitted the $H$-independent terms and $H^{\p}$-dependent ones),
$$\hat{U}_a^{(L)}=2\pi{\sinh(2\pi+\g H)k\over 2\sinh\g k
\cosh(\pi-\g)k}+\ldots.$$
Using \forphi\ we compare this expression with \repV\  (recall that
the rapidity $\alpha$ should be renormalized
$\alpha\to\t=t\alpha/2\lambda$)
and find complete agreement under the identification \param.
Similarly,
one can use
$$\hat{u}_{t-1}^{(L)}=-2\pi{\sinh(\pi+\g H)k\sinh(\pi-\g)k\over
\sinh\pi k \sinh \g k}+\ldots, $$
$$\hat{U}_{t-1}^{(L)}=-2\pi{\sinh(2+H)\g k\over  2\sinh \g k
\cosh(\pi-
\g)k}+\ldots$$
to compare $U_{t-1}$ with \repIIII\ and obtain   agreement as
well.
Next, suppose that $h>h_{th}$ ($-t<H<-1$). To obtain  the boundary
S-matrices
for scattering on the ground state $|0\>_B$ set
$\omega=\omega^{\p}=1$ and
choose the boundary string to be the longest string, $n=n_{\ast}+1$
in \physbe.
Then, using \ftII, $\hat{f}^{\p}_{n_{\ast}+1;
t-1}=-\hat{f}^{\p}_{\ast}$ and
$$\hat{u}_a^{(L)}=2\pi{\sinh\g Hk\over\sinh\pi k}+\ldots,$$
\eqn\defIII{\hat{u}_{t-1}^{(L)}=\hat{u}_a^{(L)} -
2\pi{\sinh(H+2[{1-H\over 2}])\g k \over \sinh\g k}+\ldots}
we obtain
$${\hat{U}_{n_{\ast}+1; a}^{(L)}\over 2\pi}={\sinh\g Hk\over
2\sinh\g k \cosh(\pi-\g)k}
+{\sinh(\pi-2\g)k\cosh(H+t-2n_{\ast})\g k \over  \sinh\g k
\cosh(\pi-\g)k}-$$
$$ -{2\cosh\g k \sinh n_{\ast}\g k \cosh(H-n_{\ast}+1)\g k \over
\sinh\g k \cosh(\pi-\g)k}+\dots,$$
$$\hat{U}_{n_{\ast}+1; t-1}^{(L)}=\hat{U}_{n_{\ast}+1; a}^{(L)}-
2\pi{\sinh(H+2[{1-H\over 2}])\g k\over \sinh\g k}+\ldots,$$
which agrees with \repI, \repIIII\ under the identification \param.
Note that the last relation, which follows directly from \defI,
\defII\ and
\defIII,
is valid also for  $\hat{U}_{n; a}$ and $\hat{U}_{n; t-1}$ with any
$n$.
In the same manner one can calculate the boundary S-matrices for
scattering
on the boundary n-strings  and
check that they indeed coincide   with \repII, \repIII\ under the
condition
 \param.
For this, one needs to take
$\omega=\omega^{\p}=1$ in \physbe\ and use \ftI,
$\hat{f}^{\p}_{n; t-1}=-\hat{f}^{\p}_n$. Finally one
can compute also the boundary S-matrix
for the scattering on the $(n_{\ast}+1,N)$-strings, again in agreement
with the bootstrap results.

\newsec{Conclusion}

The question of boundary bound states even in the simple  Dirichlet
case appears rather frustrating:
using the XXZ lattice regularization or equivalently
the Thirring model, we have only been able to recover
the $\beta_n$ and $\delta_{n=0,N}$ boundary bound states. A way out is
to consider solutions of the Bethe ansatz equations made of an $(n,N)$ string
superposed with the $n_{\ast}+1$ string that describes the ground state. This
is not allowed in principle in the model we consider because the Bethe
wave function vanishes
when two roots are equal. However, putting formally such a solution
in the equations gives the masses of the $\delta_{n,N}$ states and the
S-matrix also agrees with the bootstrap results! But the meaning
of this is not clear to us.

Finally let us mention that one can calculate the ground state energy
in the thermodynamic
limit  by solving the equation \beI\ for the ground state
density and using \Energy:
$$E_{gr}=\int_0^{+\infty}2L\tilde{\rho}_a(\alpha)h(\alpha)d\alpha.$$
As a result we get the combination $E_{gr}=E_{bulk}+E_{boundary}$,
where $E_{bulk}$ is the well-known sine-Gordon ground state energy
\DdVg:
$$E_{bulk}=-{Lm^2\over 4}\tan{\pi\g\over 2(\pi-\g)}$$
and $E_{boundary}$ is the contribution of the boundary terms:
$$E_{boundary}|
={m\over 2}\left[ {\sin{(H+2)\g\pi\over 2(\pi-\g)}
\over \sin{\pi^2\over 2(\pi-\g)}  } +1 + \cot{\pi^2\over 4(\pi-\g)}
\right].$$
We see that the ground state energy of the boundary sine-Gordon model
is a smooth function of the boundary magnetic field for the whole range
of $h$ in the XXZ regularization, hence of $\varphi_0$. The changes in ground
state structure do not affect $E$, as is expected since in such a unitary model
there is no
(one dimensional) boundary transition.

The finite size corrections to
the ground state energy themshelves (the genuine Casimir effect)
can be computed using the technique developed in \DdVg.
 It is also interesting to consider
the inhomogeneous 6-vertex model with an imaginary boundary magnetic field
insuring commutation with $U_q sl(2)$ \ref\PS{V. Pasquier, H. Saleur, Nucl.
Phys. B330 (1990) 523.}. This should presumably lead to a solution of minimal
models
with integrable boundary conditions.  We will report on these questions soon.

\bigskip

\leftline{\bf Acknowledgements:}
We would like to thank Paul Fendley, Anton Kapustin and
Nick Warner
for helpful discussions.
This work was supported by the Packard Foundation, the National Young
Investigator program (NSF-PHY-9357207) and  the DOE
(DE-FG03-84ER40168).

\vfill\eject

\noindent{\bf Figure Captions}
\bigskip
\noindent Figure 1: the first type of boundary string. In the ground state, the
boundary string of maximum allowed length is occupied.
\smallskip
\noindent Figure 2: the second type of boundary string.

\vfill\eject

\listrefs

 \vfill \eject
\end